# Atmospheric Influence on the Path Loss at High Frequencies for Deployment of 5G Cellular Communication Networks

Rashed Hasan Ratul
*Department of Electrical and Electronic Engineering*
*Islamic University of Technology*
Dhaka, Bangladesh
rashedhasan@iut-dhaka.edu

S M Mehedi Zaman
*Department of Electrical and Electronic Engineering*
*Islamic University of Technology*
Dhaka, Bangladesh
mehedizaman@iut-dhaka.edu

Hasib Arman Chowdhury
*Department of Electrical and Electronic Engineering*
*Islamic University of Technology*
Dhaka, Bangladesh
hasibarman@iut-dhaka.edu

Md. Zayed Hassan Sagor
*Department of Electrical and Electronic Engineering*
*Islamic University of Technology*
Dhaka, Bangladesh
zayedhassan@iut-dhaka.edu

Mohammad Tawhid Kawser
*Department of Electrical and Electronic Engineering*
*Islamic University of Technology*
Dhaka, Bangladesh
kawser@iut-dhaka.edu

Mirza Muntasir Nishat
*Department of Electrical and Electronic Engineering*
*Islamic University of Technology*
Dhaka, Bangladesh
mirzamuntasir @iut-dhaka.edu

*Abstract*— Over the past few decades, the development of cellular communication technology has spanned several generations in order to add sophisticated features in the updated versions. Moreover, different high-frequency bands are considered for advanced cellular generations. The presence of updated generations like 4G and 5G is driven by the rising demand for a greater data rate and a better experience for end users. However, because 5G-NR operates at a high frequency and has significant propagation, atmospheric fluctuations like temperature, humidity, and rain rate might result in poorer signal reception, and higher path loss effects unlike the prior generation, which employed frequencies below 6 GHz. This paper makes an attempt to provide a comparative analysis about the influence of different relative atmospheric conditions on 5G cellular communication for various operating frequencies in any urban microcell (UMi) environment maintaining the real outdoor propagation conditions. In addition, the simulation dataset based on environmental factors has been validated by the prediction of path loss using multiple regression techniques. Consequently, this study also aims to address the performance analysis of regression techniques for stable estimations of path loss at high frequencies for different atmospheric conditions for 5G mobile generations due to various possible radio link quality issues and fluctuations in different seasons in South Asia. Furthermore, in comparison to contemporary studies, the Machine Learning models have outperformed in predicting the path loss for the four seasons in South Asian regions.

I. INTRODUCTION

Millimeter-wave is a crucial component of fifth-generation technology. It is expected that 5G and B5G technology would operate at extremely high frequencies, anything from 1 GHz to 100 GHz. 5G cellular systems require enormous bandwidths that are yet to be accessible in the sub-6 GHz bands of frequencies for the purpose of fulfilling their high data throughput demands. There is a lot of excitement about the potential for 5G cellular coverage in the millimeter-wave bands, which are underutilized yet have vast accessible bandwidths. However, radio propagation at microwave and millimeter wave frequencies is fundamentally dissimilar, which has far-reaching effects on all aspects of the system, including performance and reliability and the system's response to atmospheric fluctuations [1]. In today's world, almost all cellular networks operate in the microwave spectrum. There is, thus, a lot of knowledge and data from previous measurements; nevertheless, there are significantly fewer statistical measures available for the mm-wave bands [2]. Using millimeter waves in wideband communication networks has the potential to completely revolutionize our way of life. As a result, the issue of increasing data consumption rates on mobile devices with decreasing bandwidth would be resolved [3]. Millimeter-wave telecommunications have become an integral feature of the 5G network and have laid the foundation for the next generation of communications networks. Challenges with coverage, connectivity, and quality of service (QoS) are projected to be resolved with the introduction of 5G [4]. The fifth generation mobile network, also known as 5G-New Radio (5G-NR), is a technological advancement that would systematically combine operations such as enhanced mobile broadband (eMBB), massive machine-type communication (mMTC), and ultra-reliable and low-latency communication (uRLLC), and their optimal deployment will necessitate appropriate channel modeling.

However, the millimeter waves are influenced by a wide range of atmospheric constituents, including water vapor, fog, and other impediments. This results in scattering, reflection, and diffraction, leading to a large overall impact attenuation every time the frequency climbs higher. The problem is that millimeter-wave can hardly be utilized for long-distance applications. The atmosphere influence and the propagation mechanism effect are two of the key issues in the expected reliable technological advancement in 5G and B5G. Due to atmospheric attenuation in millimeter wave networks, received signal level and path loss characteristics dynamically change depending on the combined radio link quality. As the objective of 5G communication technology is to ensure maximum reliability, the quick channel volatility under sharp atmospheric fluctuations might have a detrimental influence on availability, performance, and QoS. Thus, this article makes an attempt to investigate how various meteorological factors could influence certain mm-wave frequencies in terms of path loss performance that are specified in 3GPP Release 17 [5].

In order to carry out our investigation, four different frequencies—7.125 GHz, 24.25 GHz, 52.60 GHz, and 71 GHz—were taken into account. The simulations were carried out on the NYUSIM mm-wave channel simulator, which maintains an approximate realistic outdoor environment [6]. For the four major seasonal changes in South Asian countries, the cumulative attenuation of four key atmospheric factors like environmental temperature, effective rain rate, overall humidity, and barometric pressure, have been taken into consideration. On top of that, taking into account the overall seasonal fluctuations in some South Asian region, this research compares and contrasts the four anticipated frequency bands for 5G adoption in an urban microcell (UMi) scenario. In addition, this study can be interpreted as a pilot tutorial simulation particularly in South Asian region for evaluating the general performance characteristics of 5G installation with respect to variable weather conditions, with the aim of identifying and assessing potential atmospheric roadblocks and future opportunities in mm-wave implementations and applications. This article also demonstrates the influence of weather conditions on the path loss effects of the transmitted signal in various environments and scenarios, as well as how the millimeter-wave responds as it travels through free space and the atmosphere.

Wireless communication systems heavily rely on path loss prediction to effectively design and optimize the coverage area of a cell, which involves determining the number and placement of base stations. Accurate path loss prediction is also essential for evaluating the system's performance under various scenarios. By accurately predicting path loss, it becomes possible to optimize the transmission power and modulation scheme, leading to improved spectral efficiency, decreased interference, and extended battery life for mobile devices [7]. To enhance the accuracy of path loss prediction for 5G cellular communication deployments, several supervised machine learning algorithms have been incorporated for the selection of regression techniques. This integration of machine learning has the potential to significantly improve the accuracy of path loss prediction and optimize the performance of 5G cellular communication systems.

The remaining parts of the paper are organized accordingly: Section 2 represents the literature review. Section 3 elaborates the methodology. Section 4 describes the simulation scenario. Section 5 assesses and discusses the obtained results. Finally, section 6 draws a conclusion to the paper.

## II. LITERATURE REVIEW

At mm-wave frequencies, channel noise is relatively higher and is significantly influenced by atmospheric influences, which vary by geographical region and throughout the day. As a result, it is essential to develop accurate models for the channels of mm-wave since the wireless channel shows a considerable effect on the total system performance. In order to validate the appropriate channel model, a number of atmospheric effect studies utilizing 5G candidate frequencies have been conducted.

The influence of atmospheric pressure on a 5G channel estimation validated by codes, that used convolution techniques, was introduced in the work of Zhang et al. [7]. However, barometric pressure does not change significantly under different atmospheric conditions and might not reflect overall impact for atmospheric variations. In a recent study by Baihaqi et al. [8], the aim was to quantify the influence of both light and heavy rainfall on the performance of a model of the 5G-NR channel. The study did not account for any other types of weather influences. A study conducted by Mukhlisin et al. [9] demonstrated the impact of varying levels of humidity on the outage performance of 5G communication channels. The results provide important considerations for the design and optimization of 5G communication networks. The impact of rain on millimeter wave communication in tropical regions has been the subject of numerous studies. Two notable examples are the works by Nandi et al. [10] and Budalal et al. [11], which delve into the specifics of how rainfall can affect the performance of these systems. Their research provides valuable insights into the complex interactions between weather conditions and millimeter wave communication, and underscores the importance of accurately characterizing these effects in the design and deployment of 5G networks. However, these papers didn't consider any particular frequency bands for their research, rather, they demonstrated the impact of rainfall on a range of frequency bands. This highlights the need for further research to explore the effects of weather on specific frequency bands used in 5G networks, as these may exhibit unique performance characteristics under different environmental conditions. On the other hand, in addition to the studies previously mentioned, Kamrul et al. [12] evaluated the performance of MIMO in urban microcells in the city of Dhaka at 28 GHz, by taking into account the average weather conditions. For the 5G channel model, considering the effect of temperature in Bandung, Indonesia, the propagation characteristics at 28 GHz, 73 GHz, and 4 GHz are concentrated in the research of Rahayu et al. [13]. Though Sub- 6GHz frequencies might not be a good option for 5G deployments, analysis on the impact of rain and atmospheric gases on propagation of mm wave for 5G wireless communications were performed by Squali et al. across different seasons in various frequency bands [14].

Although the studies discussed above have examined the effects of various atmospheric impairments on mm-wave signals at different frequencies and environmental conditions, they have not taken into account the cumulative impact of all possible atmospheric conditions. Therefore, there is a need for further research that considers the combined effect of multiple atmospheric conditions on the performance of 5G networks in different regions and environments.

In this paper, a comprehensive analysis of the cumulative impact of atmospheric variations on 5G cellular communication at high frequencies was presented. The various atmospheric impairments such as atmospheric attenuation, scattering, and absorption, as well as environmental factors such as rain rate, temperature, humidity, and atmospheric pressure were considered. To assess the combined impact of these atmospheric conditions, a stability analysis is performed in the estimation of path loss using regression techniques. Multiple regression analysis is employed to develop models that relate path loss to various environmental parameters, and the models are validated using real-world measurements. The study provides innovative solutions to these issues and highlights the importance of considering weather parameters when designing and deploying 5G networks. By taking into account the impact of weather on network performance, network planners and operators can ensure that 5G networks are optimized for the local environment, resulting in improved performance and user experience. Overall, this paper provides valuable insights for researchers and industry professionals working on 5G system design and deployment.

## III. METHODOLOGY

This paper adopts two distinct approaches to analyze the impact of atmospheric conditions on the path loss of 5G cellular communication. The first strategy involves utilizing the NYUSIM package to model accurate real-world simulation scenarios, which take into account the prevailing weather conditions and overall radio link quality, to obtain simulated path loss results. To conduct a detailed evaluation of the 5G network's performance, a comprehensive simulation was carried out using real-time weather data obtained from Weather Spark, an open-source website [15]. The simulation was split into four subsections, each segment emphasizing a different season in order to achieve an in-depth examination of the network's performance under various weather conditions. This approach allows for a precise assessment of the effect of atmospheric conditions on 5G network performance.

The second strategy involves incorporating Machine Learning regression models as a validation tool for stability analysis in estimations of path loss at high frequencies for different atmospheric conditions for 5G cellular communication. This approach enables us to evaluate the accuracy of path loss predictions for various atmospheric conditions using regression analysis. By comparing the simulated path loss results with the predictions obtained from the regression models, the models to better account for the impact of different atmospheric conditions on 5G network performance can be refined. In addition, a comparison among the different regressor models was also considered to identify the best-performing model for path loss prediction under varying atmospheric conditions.

Furthermore, path loss (PL) is an essential metric to take into account while characterizing the wireless communication channel. At a carrier frequency $f_c$, in GHz, along a free space line with a benchmark distance of 1 m and attenuations due to atmospheric constraints, the following equation describes the typical form of the Cell-Individual (CI) path loss model [16].

$$\text{PL}^{\text{CI}}(f_c, d)[\text{dB}] = \text{FSPL}[\text{dB}] + 10n \log_{10}(d) + \delta[\text{dB}] + \chi_\alpha \quad (1)$$

In the above formula, the three-dimensional separation distance between transmitter end and receiver end denoted as $d$, the atmospheric attenuation is represented as $\delta$, and the path loss exponent term is specified by $n$. A random Gaussian variable $\chi_\alpha$ along with the standard deviation $\alpha$ in decibels also contributes to the overall path loss model. As seen by the aforementioned equation, distance and carrier frequency impact this model. Additionally, the equations of free space path loss (FSPL) and attenuation $\delta$ [dB] are described as follows [17]:

$$\text{FSPL}(f_c)[\text{dB}] = 20 \log_{10}\left(\frac{4\pi f_c \times 10^9}{c}\right) \quad (2)$$

$$= 32.4[\text{dB}] + 20 \log_{10} f_c$$

$$\delta[\text{dB}] = \alpha[\text{dB/m}] \times d[\text{m}] \quad (3)$$

In these preceding equations, $f_c$ stand for the carrier frequency in GHz, $c$ indicates the velocity of light, and the attenuation factor $\alpha$ is in decibels per meter in case of the frequency bands from 1-100 GHz.

## IV. SIMULATION

### A. Dataset Creation at Different Frequencies and Atmospheric Conditions using NYUSIM

The dataset was generated using NYUSIM 3.0 mm-wave software simulation to model the 5G cellular communication system in an urban microcell (UMi) environment considering Sothern Asian cities [18]. The simulation was based on real environmental data collected over the course of one year. To simplify the simulation process, four separate seasons were considered: summer, fall, winter, and spring, with each season consisting of three months as shown in Fig. 1. For each season, specific parameters like temperature, rain rate, humidity and barometric pressure were considered for four different frequencies: 7.125 GHz, 24.25 GHz, 52.60 GHz, and 71 GHz. The simulation was conducted to observe the overall path loss for a range of 10 to 500 meters for each of the four frequencies. The simulation parameters included the number and location of base stations and users, as well as the types of frequency bands used as shown in Table 1. Overall, the simulation environment was designed to accurately model the real-world conditions of an urban microcell environment in South Asian locations with typical weather, and to provide insights into the relative impact of different atmospheric conditions on 5G cellular communication for various operating frequencies.

TABLE I. OUTLINE OF SIMULATION PARAMETERS

| Input Parameters | Values/Status |
|---|---|
| Frequency (GHz) | 7.125, 24.25, 52.60, 71. |
| Channel Bandwidth | 800 MHz |
| Scenario | Urban Microcell (UMi) |
| Distance Range of Tx and Rx | 10-500 m |
| Environment | NLOS |
| Transmit Power | 30 dBm |
| Biometric Pressure (mbar) | 1000-1013 |
| Humidity (%) | 2-100 |
| Temperature | 13-40 (° C) |
| Polarization Effect | Co-polarization |
| Rain Rate ( mm/hr) | 0.2-10.5 |
| Foliage loss | No |
| Foliage Attenuation | 0.4 dB/m |
| Tx and Rx Antenna Array Type | Uniform Linear Array |
| Tx and Rx Antenna Azimuth HPBW | 10º |
| Number of Tx and Rx Antennas | 1 |
| Number of Rx Locations | 3 |
| Human Blockage Mean Attenuation | 14.4 dB |
| User Terminal Height | 1.5 m |
| Base Station Height | 32 m |

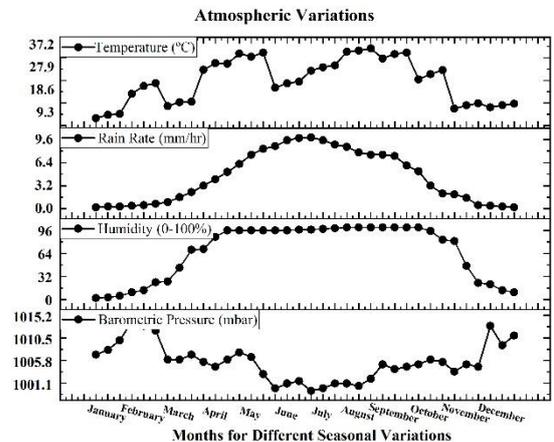

Fig. 1. Different atmospheric conditions in South Asia over a year.

## B. Comparison of Regression Technique

### 1) Dataset Description

The dataset has been produced using NYUSIM software simulations [18]. In the simulated dataset, eleven input features and one output attribute are present. The 'Data Source' and 'Simulation Number' attributes have been disregarded because they have no bearing on the 'Path Loss' output. The description of the dataset is provided in Table 2 below.

TABLE II.  OVERVIEW OF THE DATASET

| Attributes | Instances | Data Type |
|---|---|---|
| T-R Separation Distance (m) | 2835 | Continuous |
| Time Delay (ns) | | Discrete |
| Received Power (dBm) | | Continuous |
| Phase (rad) | | Continuous |
| Azimuth AoD (degree) | | Discrete |
| Elevation AoD (degree) | | Continuous |
| Azimuth AoA (degree) | | Continuous |
| Elevation AoA (degree) | | Discrete |
| RMS Delay Spread (ns) | | Continuous |
| Season | | String |
| Frequency | | Discrete |
| Path Loss (dB) | | Continuous |

### 2) Data Pre-processing:

#### a) Label Encoding

The 'Season' attribute included four seasons: 'Spring,' 'Summer,' 'Fall,' and 'Winter. Due to their string data type, it was required to convert those into numeric data. The prevalent "Label Encoder" method was used for conversion and it assigned numeric descriptors to each of the categorical values in "Season." Categorical attributes had to be converted to numeric values because ML models cannot process string-type values [19].

#### b) Supervised Models

As the output of the dataset is of continuous type, regression supervised models are to be used for training and testing the dataset [20]. For this study, nine regression algorithms have been implemented altogether and the best results among them was showcased in the results section. The research study employs a diverse set of regression techniques to predict path loss, including linear regression, robust regression, ridge regression, LASSO regression, elastic net, polynomial regression, stochastic gradient descent (SGD), random forest regressor (RF), and support vector machine regressor (SVM). These models were carefully selected based on their ability to capture the complex relationships between T-R separation distance and path loss, considering the specific characteristics of the data and the research objectives. The utilization of multiple regression techniques enhances the robustness and validity of the research findings, providing a solid foundation for the conclusions drawn from the study.

## V. RESULT ANALYSIS

### A. Software Simulation Result

The simulation findings demonstrate that the 7.125 GHz frequency exhibited a relatively stable path loss effect across all four seasons, while there was a consistent increase in path loss as the Transmitter-Receiver (T-R) separation distance was expanded. This trend can be attributed to the longer wavelength of 7.125 GHz, which enables it to effectively penetrate obstacles and travel further distances. Conversely, the higher frequencies, namely 24.25 GHz, 52.60 GHz, and 71 GHz, displayed a relatively higher path loss level, regardless of the TR separation distance. As the TR separation distance was further extended, the path loss for higher frequencies also increased marginally. This phenomenon is due to the shorter wavelength of higher frequencies, which makes them more prone to atmospheric absorption and scattering. It is noteworthy that no significant variations were observed in path loss effects across the four seasons for all frequencies under examination. This finding suggests that atmospheric conditions, such as rain rate, humidity, and temperature, did not have a marked impact on path loss for the frequencies investigated in this study.

Finally, the 7.125 GHz frequency performed well in closer distances where the transmitter and receiver are located nearby. However, at greater separation distances, the 7.125 GHz frequency performed better than the other frequencies, implying that it could be a more suitable option for long-range communication in urban microcell environments. The following four figures (Fig. 2, Fig. 3, Fig. 4, Fig. 5) presented in this research paper depict the path loss versus T-R separation distance under four different seasons. These figures provide visual representations of the impact of T-R separation distance on path loss for the frequencies under investigation in various seasons, including fall, winter, summer, spring.

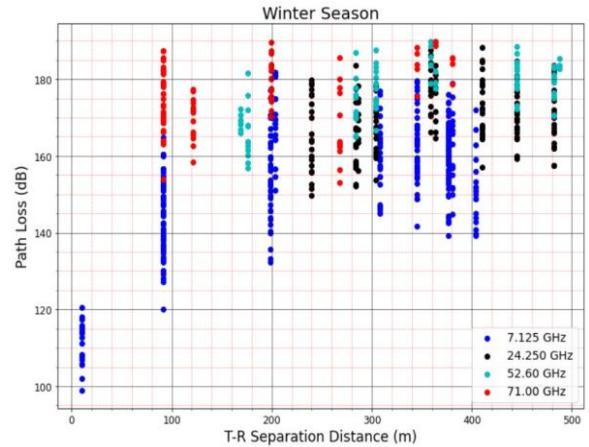

Fig. 2.  Path loss vs. T-R separation in winter

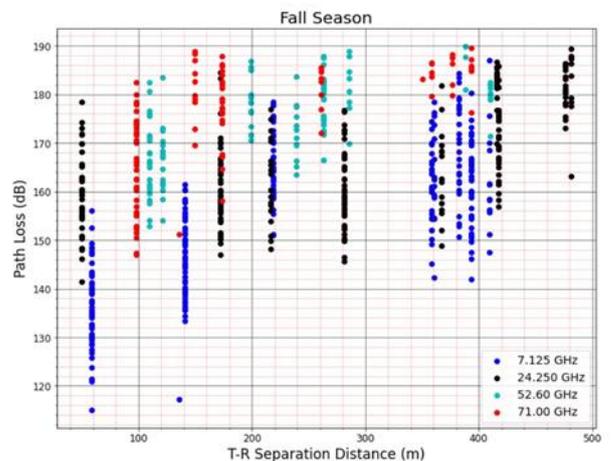

Fig. 3.  Path loss vs. T-R separation in fall

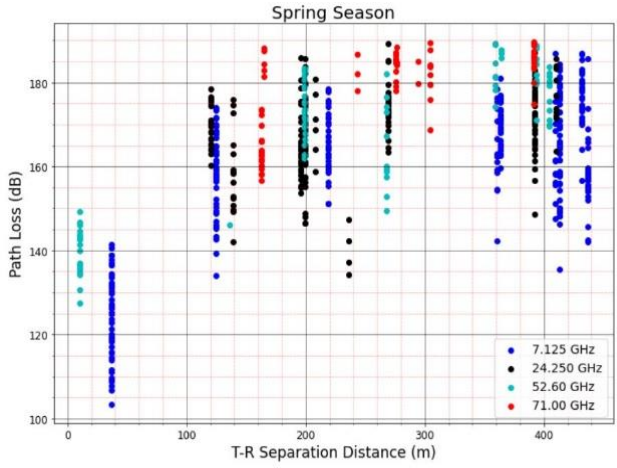

Fig. 4. Path loss vs. T-R separation in spring

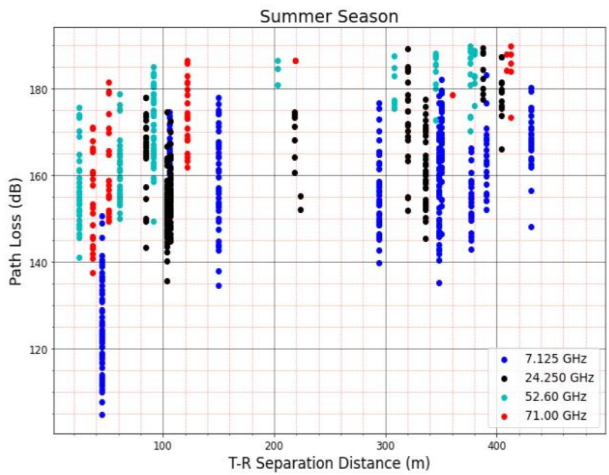

Fig. 5. Path loss vs. T-R separation in summer

*B. Path Loss Prediction Result using ML*

The research study evaluates the performance of the regression models using various prediction metrics, including Mean Absolute Error (MAE), Mean Squared Error (MSE), Root Mean Squared Error (RMSE), and R-squared (R2). The results are presented in Table 3, and it is evident that the Random Forest Regressor outperforms the other models for this particular dataset.

TABLE III. PERFORMANCE ANALYSIS

| Models | MAE | MSE | RMSE | R2 |
|---|---|---|---|---|
| Linear Regression | 5.061 | 42.777 | 6.540 | 0.813 |
| Robust Regression | 5.596 | 70.153 | 8.375 | 0.693 |
| Ridge Regression | 5.506 | 48.404 | 6.957 | 0.788 |
| LASSO Regression | 7.827 | 97.061 | 9.851 | 0.576 |
| Elastic Net | 5.202 | 43.594 | 6.602 | 0.809 |
| Polynomial Regression | 4.388 | 33.833 | 5.816 | 0.852 |
| SGD | 5.414 | 46.738 | 6.836 | 0.796 |
| **RF Regressor** | **3.485** | **24.809** | **4.980** | **0.891** |
| SVM Regressor | 6.687 | 82.902 | 9.105 | 0.638 |

R2 metric, as shown in Fig 6, clearly indicates the superior performance of the Random Forest Regressor. These findings highlight the effectiveness and accuracy of the Random Forest Regressor in predicting path loss in the specific research context. The results provide valuable insights for the selection of the most appropriate regression model for future studies and practical applications in similar scenarios.

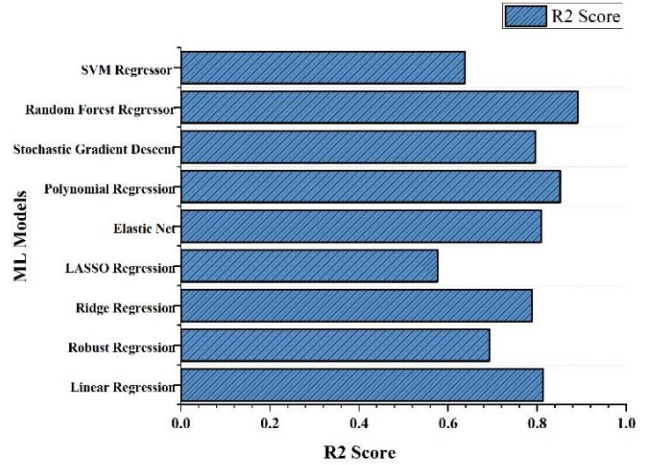

Fig. 6. Comparison of R2 performance metric among all nine algorithms

Table 4 presents a comprehensive comparison of the performance metrics obtained in this study with those reported in previous path loss prediction studies. The findings highlight the superiority of the regression models used in this study, as evidenced by the lower values of MAE, MSE, and RMSE, and the higher value of R2. These results indicate that the proposed models in the present study outperform the models used in previous studies in terms of predictive performance.

TABLE IV. COMPARATIVE ANALYSIS WITH PREVIOUS STUDIES

| References | MAE | MSE | RMSE | R2 |
|---|---|---|---|---|
| [21] | 4.74 | 39.38 | 6.27 | - |
| [22] | - | - | 8.67 | - |
| [23] | 4.28 | - | 5.60 | - |
| [24] | 5.10 | 44.51 | 6.67 | 0.72 |
| **This Study** | **3.485** | **24.809** | **4.980** | **0.891** |

Fig. 7 presents a graphical comparison of the Root Mean Squared Error (RMSE) performance obtained in the current research study with that reported in other relevant articles on path loss prediction. The superior performance of the regression models utilized in the study is clearly illustrated by the lower RMSE values observed as compared to the values reported in other articles. The lower RMSE values reflect better accuracy and predictive performance of the proposed models in estimating path loss in the specific research context. This comparison substantiates the effectiveness of the selected regression techniques in this study and further reinforces the reliability and robustness of the findings presented.

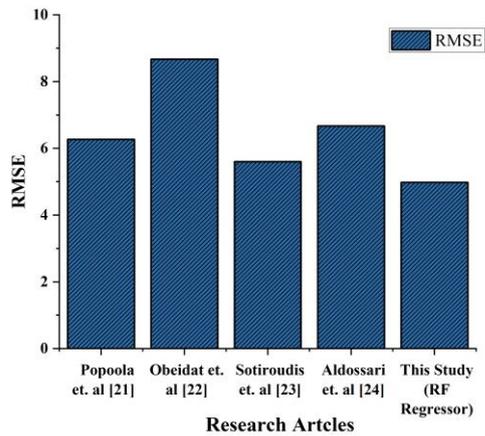

Fig. 7. Comparison of RMSE performance with contemporary articles

## VI. Conclusion

The findings of this research paper highlight the importance of frequency selection in wireless communication systems, particularly in urban microcell environments. The simulation results reveal that the 7.125 GHz frequency exhibits relatively stable path loss characteristics across all seasons and performs well in both short and long-range communication scenarios. The use of ML models has been found to have a significant impact on accurately predicting path loss for the frequencies under investigation. Furthermore, the lack of significant variations in path loss effects across different seasons suggests that atmospheric conditions have minimal impact on the frequencies examined in this study. Further research and experimentation in real-world environments are warranted to validate and extend these findings, and to optimize the performance of wireless communication systems in urban microcell environments.